\documentstyle[prl,aps,twocolumn,epsf]{revtex}  
  
\catcode`\@=11

\def\maketitle2{\par 
\begingroup
\let\cite\@bylinecite
\def\thefootnote{\fnsymbol{footnote}}%
\twocolumn[\@maketitle2\vskip2pc]%
\thispagestyle{plain}\@thanks
\endgroup
\def\thefootnote{\arabic{footnote}}%
\setcounter{footnote}{0}%
\let\maketitle2\relax \let\@maketitle2\relax
\let\@thanks\relax \let\@authoraddress\relax \let\@title\relax
\let\@date\relax \let\thanks\relax \let\@abstract\relax 
\let\@pacs\relax}

\def\abstract#1{\gdef\@abstract{{\par 
\bgroup
\ifdim\prevdepth=-1000pt \prevdepth0pt\fi
\hsize\columnwidth
\dimen0=-\prevdepth \advance\dimen0 by17.5pt \nointerlineskip
\small\vrule width 0pt height\dimen0 \relax}{~~}#1\egroup}}

\def\pacs#1{\gdef\@pacs{{\par 
\bgroup
\hsize\columnwidth \parindent0pt
\ifdim\prevdepth=-1000pt \prevdepth0pt\fi
\dimen0=-\prevdepth \advance\dimen0 by20pt\nointerlineskip
\egroup} PACS numbers:~#1}}

\def\@maketitle2{
\@preprint
\@title
\ifdim\prevdepth=-1000pt \prevdepth0pt\fi
\@authoraddress
\@date
\begin{list}{}{\leftmargin=0.10753\textwidth \rightmargin=\leftmargin
\itemsep=1pc\partopsep=-1pc}
\item\@abstract
\item\@pacs
\end{list}
}

\catcode`\@=12

\makeatletter  
  
\def\compoundrel#1\over#2{\mathpalette\compoundreL{{#1}\over{#2}}}  
\def\compoundreL#1#2{\compoundREL#1#2}  
\def\compoundREL#1#2\over#3{\mathrel  
  {\vcenter{\hbox{$\m@th\buildrel{#1#2}\over{#1#3}$}}}}  
\makeatother  
  
\begin{document}  
  
\draft  
  
\title{Discrete Wigner functions and the phase space representation of 
quantum computers}

\author{Pablo Bianucci$^1$, Cesar Miquel$^1$, Juan Pablo Paz$^1$ and Marcos  
Saraceno$^2$}  
  
\address{$^1$Departamento de F\'{\i}sica ``J.J. Giambiagi'',  
FCEN, UBA, Pabell\'on 1, Ciudad Universitaria, 1428 Buenos Aires, Argentina}  
  
\address{$^2$Unidad de Actividad F\'{\i}sica, Tandar, CNEA  
Buenos Aires, Argentina}  
  
  
\abstract  
{We show how to represent the state and the evolution of a quantum computer
(or any system with an $N$--dimensional Hilbert space) in phase  
space. For this purpose we use a discrete version of the Wigner function  
which, for arbitrary $N$, is defined in a phase space grid of $2N\times 2N$  
points. We compute such Wigner function for states which are relevant 
for quantum computation. Finally, we discuss properties of quantum 
algorithms in phase space and present the phase space 
representation of Grover's quantum search algorithm.}

\date{\today}  
  
%
%

\maketitle2  
\narrowtext  

Wigner functions \cite{wigner} provide a simple representation of 
the quantum state of a continuous system in phase
space. This is important when analyzing issues  
related with the classical limit of quantum mechanics
\cite{decowigner}. In this letter  
we study a useful generalization of the Wigner function  
to systems with $N$--dimensional Hilbert spaces. We  
use this Wigner function to obtain, for the first time, a   
phase space representation of both the states and the temporal  
evolution of a quantum computer.
What are the potential advantages of a phase space representation
of a quantum computer? It is clear that quantum
algorithms can be thought of as quantum maps operating in a
finite state space and are therefore
amenable to this representation. Whether it will be useful or not will
depend on properties of the algorithm.
Specifically, algorithms become interesting in the large $N$ limit 
({\it i.e.} when operating on many qubits). For a quantum map
this is the {\it semiclassical limit} where many new regularities arise in
connection with its classical behaviour in phase space. Thus unraveling
these regularities, when they exist, becomes an important issue which 
can be naturally accomplished in a
phase space representation. Moreover, this phase space 
approach may allow one to establish contact between the vast literature on 
quantum maps (dealing with their construction, semiclassical properties, etc) 
and that of quantum algorithms providing hints to develop new
algorithms. As a first application of these ideas 
we show how Grover's search algorithm \cite{grover} can be represented 
in phase space and interpreted as a simple quantum map.

For a particle in $1$ dimension the Wigner function, \cite{wigner}  
\begin{equation}  
W(q,p)=\int {d\lambda\over{2\pi\hbar}}  
e^{-i\lambda p/\hbar}    
\langle q-{\lambda\over 2} |\hat\rho |q+{\lambda\over 2}  
\rangle,\label{wigcont}.   
\end{equation}  
has the following properties:   
(P1) $W(q,p)$ is real valued,   
(P2) The inner product between states $\rho_1$ and $\rho_2$ is   
$Tr(\rho_1\rho_2)=2\pi\hbar\int dqdp W_1(q,p)W_2(q,p)$,
(P3) The integral of $W(q,p)$ along a line in phase space,     
defined as $a_1 q+a_2 p=a_3$, is the probability density  
that a measurement of the observable $a_1\hat Q+a_2\hat P$ has $a_3$   
as a result. These properties follow directly  
from the definition (\ref{wigcont}). However, their origin can 
be better understood by noticing that $W(q,p)$ is the expectation value of  
an operator $\hat A(q,p)$, known as ''phase space point  
operator":
\begin{equation}  
W(q,p)=Tr(\hat\rho \hat A(q,p)).\label{wig1}  
\end{equation}  
The operator $\hat A(q,p)$, that parametrically
depends on $(q,p)$, is a symmetrized product of delta functions:
\begin{equation}  
\hat A(q,p)=\int\!\!\int {d\lambda d\lambda'\over{(2\pi\hbar)^2}}\  
e^{i{\lambda\over\hbar} (\hat Q-q) - i{\lambda'\over\hbar} (\hat P-p)} 
\label{apqcont}  
\end{equation}  
Properties (P1-P3) follow directly from simple   
properties of $\hat A(q,p)$. In fact, reality  
of $W(q,p)$ is a consequence of the hermiticity of $\hat A(q,p)$. 
The inner product rule follows from the fact  
that $\hat A(q,p)$ are complete in the sense that $Tr(\hat A(q,p)  
\hat A(p',q'))=\delta(q-q')\delta(p-p')/2\pi\hbar$. Finally, property (P3) is
a consequence of the fact that when integrating 
$\hat A(q,p)$ over a line in phase space one gets a   
projection operator: $\int dp dq \delta(a_1q+a_2 p-a_3)\hat A(q,p)=  
|a_3\rangle\langle a_3|$ where $|a_3\rangle$ is an eigenstate 
of the operator   
$a_1\hat Q+a_2 \hat P$ with eigenvalue $a_3$. Notice that although the
above definitions are given in terms of $\hat P$ and $\hat Q$,
only unitary exponentials of such operators (i.e, finite phase space translations) 
are involved in (\ref{apqcont}).
 
To define Wigner functions for discrete systems various attempts can be  
found in the literature. Most  
notably, Wooters \cite{Wooters} proposes a definition that has the   
properties (P1--P3) only when $N$ is a prime number. His phase space
is an $N\times N$ grid (if $N$ is prime) and a cartesian product of   
spaces corresponding to prime factors of $N$ in the general case.   
Rivas and Ozorio de Almeida \cite{Rivas} define translation   
and reflection operators relating to the geometry of chords and centers on the
phase space torus. Bouzouina and De Bievre \cite{Bouzouina}   
give a more abstract derivation relating to   
geometric quantization. Our approach here follows closely Wooters   
ideas but combine them with the results in \cite{Rivas} to   
obtain a Wigner function with all the required properties (including the  
generalization of (P3)).

Given a basis of Hilbert space (for example, the computational 
basis of a quantum
computer) $B_x=\{|n\rangle, n=0,..,N-1\}$ we can think of it as a
discretized coordinate basis (with periodic boundary conditions).
Its discrete Fourier transform $B_p=\{|k\rangle, k=0,..,N-1\}$, 
related to $B_x$ by
$|k\rangle=\sum_n \exp(i2\pi nk/N)|n\rangle/\sqrt N$, can be considered 
as the corresponding momentum basis (with the same periodicity).
For a system with an $N$ dimensional Hilbert space it is not possible
to define hermitian position and momentum operators that generate infinitesimal
displacements. As Schwinger has shown \cite{Schwinger}, this shortcoming is 
overcome by introducing {\it unitary} operators that generate 
finite displacements.
The translation operator $\hat U$ generates cyclic shifts in the 
position basis, {\it i.e.}
$\hat U^m|n\rangle = | n+m\rangle $. On the other hand, displacement operators
in the momentum basis satisfy $\hat V^m|k\rangle = |k+m\rangle $
and are diagonal in the $B_x$ basis:
$\hat V^m | n \rangle = \exp(i2\pi mn/N)|n\rangle$ 
(analogously,
$\hat U$ is diagonal in the $B_p$ basis: 
$\hat U^n | k \rangle = \exp(-i2\pi kn/N)|k\rangle$).
The commutation relations between $\hat U$ and $\hat V$ take the form
$\hat U \hat V= \hat V \hat U \exp{(-2i\pi/N)}$ \cite{Schwinger}.
Using this it is easy to show that the operator
$\hat T(m,k)=\hat U^m \hat V^k \exp(i\pi mk/N)$ is the discrete 
equivalent of the
symmetrized displacement $\exp{{i\over\hbar}( q \hat P - p \hat Q)}$ in the
continuous case.

With these tools, it is natural to generalize (\ref{apqcont}) as
$$
\hat{\cal A}(q,p)={1\over N^2}\sum_{m,k=0}^{N-1}
\hat T(m,k)\exp{\{-i2\pi{(kq-mp)\over N}\}}.
$$
However, one immediately sees that this definition has a serious problem: Thus,
for real values of $(q,p)$ the operators defined in this way are not hermitian.
This problem comes as no surprise since the use of an integer grid does
not allow the half-points needed in (\ref{wigcont}).
As previous works indicate \cite{Hannay}, there is an obvious
solution to this problem which simply consists of defining Wigner   
functions in an array of $2N\times 2N$ points
(lying at half-integer values of $q,p$).
This leads to the following definition of phase space point operators
in the discrete case
\begin{equation}
\hat A(q,p)={1\over (2N)^2}\sum_{\lambda,\lambda'=0}^{2N-1} 
\hat T(\lambda,\lambda')\exp{\{-i2\pi{(\lambda' q -\lambda p)\over 2N}\}}
\label{disphase}
\end{equation}
where now $q,p$ are integers in the range $\{0,...,2N-1\}$

It is simple to show that these operators have all the desired properties
and enable us to define the Wigner function exactly as in (\ref{wig1})   
(i.e., the discrete Wigner function is the expectation value of the discrete  
phase space point operator (\ref{disphase})).
First, the operators $\hat A(q,p)$ are Hermitian.
Second, they form a complete set. Therefore, one can always invert  
equation (\ref{wig1}) and express $\hat\rho$  
as a linear combination of $\hat A(q,p)$. In fact,   
the Wigner function are the coefficients of such expansion: i.e.  
\begin{equation}
\hat\rho=N\sum_{p,q}^{2N-1} W(q,p) \hat A(q,p).\nonumber
\end{equation}    
Using this, it is simple to show that the inner product between two   
states is obtained from the corresponding Wigner functions   
as: $Tr(\rho_1\rho_2)=N\sum_{q,p}W_1(q,p)
W_2(q,p)$. It is worth noticing that the number of phase space point  
operators is $4N^2$ but that there are only $N^2$ of them that are
linearly independent. A basis set for such operators is formed by
the ones corresponding to points with $0\le q,p\le N-1$, belonging to   
the first square subgrid of $N\times N$ points of the original
$2N\times 2N$ lattice. Thus, the entire Wigner function is
determined by the values it takes on this first square subgrid.
In fact, one can show that
$W(\sigma_q N+q,\sigma_p N+p)=(-1)^{\sigma_q p+\sigma_p q + N\sigma_p\sigma_q} W(q,p)$,
an expression that relates the values of the Wigner function on the
four different $N\times N$ subgrids.
  
Generalizing property (P3) to the discrete case is 
less trivial. Following Wooters \cite{Wooters}, we can define    
a line $L$ in the phase space grid as the set of points
$L(n_1,n_2,n_3)=\{(q,p) {\rm\ such\ that\ } n_1 q+n_2 p=n_3\}$
with $0\le n_i\le 2N-1$ (two lines are parallel if they are   
parametrized by the same integers $n_1$ and $n_2$ but differ in $n_3$).  
It is possible to show that the above Wigner function  
satisfies the following crucial property:
When adding its value over all points belonging to a 
line $L(n_1,n_2,n_3)$ one always gets   
a positive number which is nothing but the probability of measuring  
the eigenstate with eigenvalue $\exp(i2\pi n_3/N)$ of the operator   
$\hat T(n_1,n_2)$
(if $\hat T(n_1,n_2)$ does not have $n_3$ as one of its  
eigenphases, the sum of $W(q,p)$ over $L(n_1,n_2,n_3)$ is equal to zero).   
This is a consequence of the fact that  
when adding all phase space point operators in   
the line $L(n_1,n_2,n_3)$ one gets a projection operator. Thus,   
$$\hat A_L=  
\sum_{(q,p)\in L}\hat A(q,p)={1\over N}\sum_\nu\delta(2\nu-n_3)|\nu\rangle  
\langle\nu|$$ where the states $|\nu\rangle$ are eigenstates of 
$\hat T_{n_1,n_2}$  
with eigenvalue $\exp(-i2\pi\nu/N)$. 
We have shown that (for even values of $N$) the operator $\hat A_L$ 
projects onto a subspace whose dimensionality is equal 
to $d_L=D_L/N$ where $D_L$ is the number of points in the 
line $L$ for which $p$ and $q$ are even (clearly this implies, 
for example, that $D_L=0$ if $n_3$ is odd). 
As an illustration, let  
us apply this to the simplest case: For a line $L_q$ defined  
as $q=n_3$, the Wigner function summed over   
all points in $L_q$ is $\sum_{(q,p)\in L_q} W(q,p)=\sum_p W(q,p)=  
\langle q/2|\hat\rho|q/2\rangle$ if $q$ is even  
(and zero otherwise). Analogously, considering horizontal lines
($L_p$ defined as $p=n_3$) we get  
$\sum_{(q,p)\in L_p} W(q,p)=\sum_q W(q,p)=  
\langle p/2|\hat\rho|p/2\rangle$ if $p$ is even  
(and zero otherwise). These are just two examples a general result: 
this Wigner function always generates the correct marginal 
distributions (this, as in the  
continuous case, is the defining feature of $W(q,p)$).  
  
Now, let us discuss the properties of the Wigner function of    
states of a quantum computer. If we consider a computational state   
with $\hat\rho=|n_0\rangle\langle n_0|$ (a position eigenstate), it is   
simple to show that 
$W(q,p)=\delta_N(q-2n_0) (-1)^{p(q-2n_0)_N}/2N$ (where $z_N$ 
denotes $z$ modulo $N$). This function, displayed in Figure 1, 
is nonvanishing only along two vertical strips located at 
$q=2n_0$ modulo $N$. In one of these two strips ($q=2n_0$)  
$W(q,p)$ is positive while in the other one ($q=2n_0\pm N$)   
it oscillates becoming negative  
in odd values of $p$. These oscillations are   
typical of interference patterns and, in fact, can be interpreted  
here as originating from the interference between the positive strip and  
its mirror image created by the periodic boundary conditions we  
are imposing by requiring ciclic behavior modulo $N$. The fact that the  
Wigner function becomes negative in this interference strip is essential   
to recover the correct marginal distributions: In fact, adding values of  
$W(q,p)$ along vertical lines
gives the probability of measuring a computational state, which
is zero for all states and $1$ for state $|n_0\rangle$. The Wigner function  
of a momentum eigenstate is entirely analogous to the one shown in Figure 1a  
but rotated $90$ degrees.   
Another interesting state is shown in Figure 1b where one sees the   
Wigner function of a superposition of two computational states. A closed  
analitic expression is easy to write down but the graphic representation  
is much more eloquent: For the quantum state $|\Psi\rangle  
=(|n_0\rangle+|n_1\rangle)/\sqrt 2$, $W(q,p)$ is non zero on the two   
positive strips located at $q=2n_0$ and $q=2n_1$ and also on a strip   
located in between these two where it oscillates due to interference.   
The wavelength of these fringes depends on the distance between the   
two interfering states and is given by $2N/(n_1-n_0)$). On the other hand  
$W(q,p)$ is also nonzero and oscillatory on the strips obtained from   
the interference between the above three strips and their corresponding   
mirror images. Other states have simple Wigner functions:
for the identity, $W(q,p)$ is zero everywhere except when $p$ and $q$ 
are both even where it is equal to $1/N$.

Temporal evolution of a discrete quantum system has also a natural 
phase space representation. In fact, the evolution of $\rho$ as 
$\rho'={\cal U}\rho{\cal U^\dagger}$ implies that the Wigner 
function transforms as $W'(\alpha)=\sum_\alpha Z_{\alpha\beta}W(\beta)$, 
where 
$Z_{\alpha\beta}=Tr(\hat A_\alpha{\cal U} \hat A_\beta {\cal U^\dagger})$. 
The unitarity of ${\cal U}$ implies that $Z_{\alpha\beta}$ 
is real and orthogonal (here $\alpha$ and $\beta$ denote points in 
the first $N\times N$ phase space subgrid, i.e. 
$\alpha=(q_\alpha,p_\alpha)$ with $0\le q_\alpha, p_\alpha <N$).  
Analyzing quantum evolution in phase space is convenient if one
wants to reveal properties of the semiclassical (large $N$) regime. 
The criterion for an evolution operator ${\cal U}$ to have a 
classical analog is simple: This happens if the matrix $Z_{\alpha\beta}$ 
defines a deterministic map in phase space. This is so if, for every
point $\beta$ the matrix element $Z_{\alpha\beta}$ is nonzero only if 
$\beta=M(\alpha)$ (in this case, the present value of the Wigner function in 
the point $\alpha$ uniquely determines the future value the Wigner function
at the point $\beta=M(\alpha)$), This is precisely what 
happens in a classical dynamical flow. In such case we can say that  
the quantum evolution is simply associated with the classical map $M$ (i.e., 
$W(\alpha)=W'(M(\alpha))$). 
Finding unitary operators with this property is simple (but, clearly, generic
operators do not satisfy this criterion). In fact, operators proportional to 
all phase space point operators $\hat A(\alpha)$ are classical in 
that sense as well as all phase space translations $\hat T(m,k)$ and 
all unitary operators associated with linear canonical 
transformations (quantum cat maps \cite{Hannay}). 
Thus, it is simple to show that the matrix $Z_{\alpha\beta}$ associated 
with ${\cal U}=2N \hat A(\alpha_0)$ is 
$Z_{\alpha\beta}=\delta_N(\alpha+\beta-2\alpha_0)$ (corresponding to a 
classical map obtained by a reflection followed by a phase space translation). 
On the other hand, for ${\cal U}=\hat T(m,k)$ we 
have $Z_{\alpha\beta}=\delta_N(\alpha-\beta - 2\alpha_0)$ 
(which is simply a phase space translation by $\alpha_0=(m,k)$). 
Notice that the factor of $2$ in the above translations is a clear 
indication of the need for introducing a phase space grid of 
size $2N\times 2N$. 
To the contrary, nonlocal operations are very common: Products of 
Pauli operators acting on different qbits are nonlocal in phase space. 
For example, the matrix corresponding to ${\cal U}=\sigma_z^{(0)}$ 
(where the Pauli operator acts on the least significant qbit) is such 
that $Z_{0\beta}=2/N$ for even values of $\alpha$ being equal to zero 
otherwise. 

\begin{figure}
\epsfxsize=8.6cm
\epsfbox{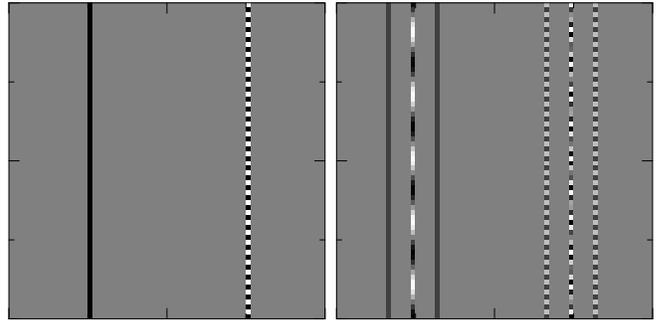}
\vspace {0.25cm}
\caption{Left: Wigner function of a computational state of $5$ qubits
--black (white) correspond to positive (negative) values--.
The oscillating strip arises from the periodic boundary conditions.
Right: Wigner function for a superposition of two computational states
has two positive strips, an intermediate oscillating one and three other
strips arising from the interference between the above and their mirror
images.}
\label{figure1}
\end{figure}

Existing quantum algorithms are a complicated combination of classical 
and quantum operations that, in general, may not have a simple 
phase space representation. However, there are remarkable 
exceptions: For example, the discrete Fourier transform ${\cal U}_{FT}$, 
which plays a major role in quantum algorithms \cite{grover},  
has a very simple phase space representation (this is not surprising 
since the very notion of conjugate variables relies on ${\cal U}_{FT}$). 
In fact, the Fourier transform is represented in phase space by a dynamical 
map whose action is $W'(q,p)=W(-p,q)$ (the Wigner function of the state
after applying ${\cal U}_{FT}$ is obtained from the original one by 
changing the phase space point $(q,p)$ into $(-p,q)$, a canonical transformation
that corresponds to a $\pi/2$--rotation in phase space).
Another important 
example of a quantum algorithm with a simple phase 
space representation is Grover's quantum search whose phase space action
is shown in Figure 3. In such Figure we display the Wigner function of the   
quantum state of a computer after every iteration of Grover's search  
algorithm. Our system has an $N=32$--dimensional Hilbert space (i.e.,   
the computer has just $5$ qubits) and the algorithm is designed to   
search for the marked item which we chose here to be $q=16$ (indicated with   
an arrow in the plot). The initial state is chosen to be an equally   
weighted superposition of all computational states which, for the  
purpose of making the plots more visible, we chose to be a non-zero  
momentum state (the usual choice for the initial state  
in Grover's algorithm is a zero  
momentum state but the algorithm works as well with an initial state  
with nonzero momentum).

\begin{figure}
\epsfxsize=8.6cm
\epsfbox{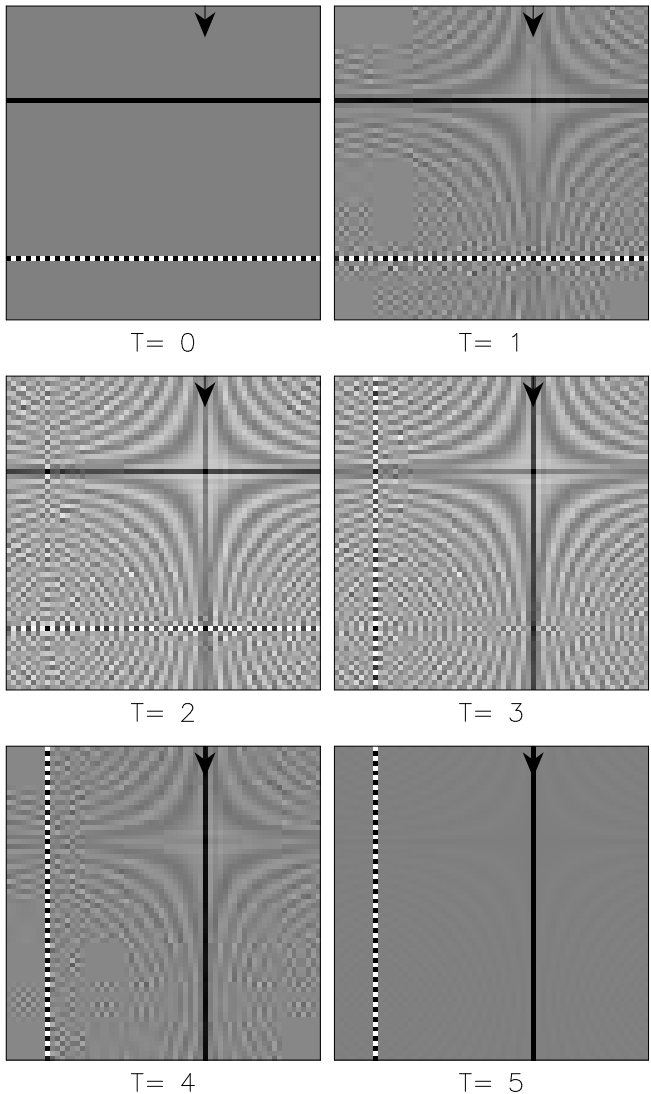}
\vspace{0.25cm}
\caption{Phase space representation of Grover's algorithm. The computer
starts in a pure momentum state and evolves (after only five iterations)
into a position eigenstate corresponding to the searched item.}
\label{figure3}
\end{figure}

One clearly observes that the algorithm is very  
simple when seen from a phase space representation. The initial state   
has a Wigner function which is a horizontal strip (with its   
oscillatory companion). After each iteration $W(q,p)$ shows a
very simple Fourier-like pattern and becomes a pure coordinate state at the
end of the search (in our case, the optimal number of iterations 
is $T\approx\pi\sqrt N/4\approx 5$).
This representation shows that, as a map, Grover's algorithm
has a fixed phase space point with coordinate equal to the marked item ($q=16$
in our case) and momentum equal to the one of the initial state.

We have presented a method to represent both the states and the evolution 
of a finite quantum system
in phase space. The method has points of contact, and also differences,
with previous \cite{Rivas,Bouzouina,Hannay,Galetti} and more recent approaches
\cite{recent}. The main
difference that arises with respect to the usual continuous infinite dimensional
case is the fact that finite dimensionality (and periodic boundary conditions
implicit in modulo $N$ arithmetic) produces a characteristic interference pattern
between a ``fundamental'' Wigner function and its periodized images 
(clearly illustrated in the Figures). Such fringes, that can also be
interpreted as arising from the characteristic {\it aliasing} implicit in the
discrete Fourier transform, are absolutely essential to provide the correct
marginal distributions from the Wigner function.
Finally, regarding potential
applications to study quantum algorithms, we believe that this method provides
a novel way to analyze their behaviour both in the computational and the complementary
Fourier transformed basis, thus paving the way for a semiclassical analysis of
algorithms.
This work was supported by grants from Anpcyt (PICT 01014), Ubacyt and Conicet.

%
%

%
%

\end{document}